\documentclass[sigconf]{acmart}

\usepackage{booktabs}
\usepackage{todonotes}
\usepackage[ruled,vlined]{algorithm2e}

\AtBeginDocument{%
  \providecommand\BibTeX{{%
    \normalfont B\kern-0.5em{\scshape i\kern-0.25em b}\kern-0.8em\TeX}}}

\copyrightyear{2020}
\acmYear{2020}
\setcopyright{acmcopyright}
\acmConference[CIKM '20] {The 29th ACM International Conference on Information and Knowledge Management}{October 19--23, 2020}{Virtual Event, Ireland}
\acmBooktitle{The 29th ACM International Conference on Information and Knowledge Management (CIKM '20), October 19--23, 2020, Virtual Event, Ireland}
\acmPrice{15.00}
\acmDOI{10.1145/3340531.3412046}
\acmISBN{978-1-4503-6859-9/20/10}

\theoremstyle{definition}
\newtheorem{definition}{Definition}[section]
\theoremstyle{hypothesis}

\begin{document}
\fancyhead{}

\title{FANG: Leveraging Social Context for Fake News Detection Using Graph Representation}


\author{Van-Hoang Nguyen}
\email{vhnguyen@u.nus.edu}
\affiliation{
  \institution{National University of Singapore}
  \streetaddress{13 Computing Drive}
   \state{Singapore}
  \postcode{117417}
}

\author{Kazunari Sugiyama}
\email{kaz.sugiyama@i.kyoto-u.ac.jp}
\affiliation{
  \institution{Kyoto University}
  \streetaddress{Yoshida-honmachi, Sakyo-ku}
  \city{Kyoto}
   \state{Japan}
  \postcode{606-8501, Japan}
}

\author{Preslav Nakov}
\email{pnakov@hbku.edu.qa}
\affiliation{
  \institution{Qatar Computing Research Institute, HBKU}
  \streetaddress{HBKU Research Complex, RC1}
    \city{Doha}
   \state{Qatar}
  \postcode{P.O. box 34110}
}

\author{Min-Yen Kan}
\email{kanmy@comp.nus.edu.sg}
\affiliation{
  \institution{National University of Singapore}
  \streetaddress{13 Computing Drive}
   \state{Singapore}
  \postcode{117417}
}

\renewcommand{\shortauthors}{Nguyen, Sugiyama, Nakov, and Kan}

\begin{abstract}
We propose Factual News Graph (FANG), a novel graphical social context representation and learning framework for fake news detection. Unlike previous contextual models that have targeted performance, our focus is on representation learning. Compared to transductive models, FANG is scalable in training as it does not have to maintain all nodes, and it is efficient at inference time, without the need to re-process the entire graph. Our experimental results show that FANG is better at capturing the social context into a high fidelity representation, compared to recent graphical and non-graphical models. In particular, FANG yields significant improvements for the task of fake news detection, and it is robust in the case of limited training data. We further demonstrate that the representations learned by FANG generalize to related tasks, such as predicting the factuality of reporting of a news medium. 
\end{abstract}




\maketitle

\section{Introduction}\label{sec:introduction}

\begin{table*}[tbh]

\caption{Engagement of social media users with respect to fake and real news articles. Column 2 shows the time since publication, and columns 4--7 show the distribution of stances  (S: Support, D: Deny, C: Comment, and R: Report).
}
  \label{table:temporal_engagement}
  \centering
  \small
  \begin{tabular}{p{4cm}ccccccp{6.5cm}}
    \toprule
    \bf News title \textbf{(Label)} & \bf Time & \bf \# Posts & \bf S & \bf D & \bf C & \bf R & \bf Noticeable responses \\ 
    \midrule
  Virginia Republican Wants Schools To Check Children's Genitals & 3h & 38 & 0.00 & 0.03 & 0.19 & 0.78 & ``DISGUSED SO TRASNPHOBIC'', ``FOR GODS SAKE  GET REAL GOP'', ``You cant make this up folks'' \\ \cline{2-8}
  Before Using Bathroom \textbf{(Fake)}  & 3h - 6h & 21  & 0.00 & 0.10 & 0.10 & 0.80 & ``Ok This cant be real'', ``WTF IS THIS BS'', ``Rediculous RT'' \\ \cline{2-8}
  & 6h+ & 31 & 0.00 & 0.10 & 0.14 & 0.76 & ``Cant make this shit up'', ``how is this real'', ``small government'', ``GOP Cray Cray  Occupy Democrats'' \\ \hline
  1,100,000 people have been killed by & 3h & 9 & 0.56 & 0.00 & 0.00 & 0.44 & ``\#StopGunViolence'', ``guns r the problem'' \\ \cline{2-8}
  guns in the U.S.A. since John Lennon was shot and killed on December 8, 1980 \textbf{(Real)} & 3h+ & 36 & 0.50 & 0.00 & 0.11 & 0.39 & ``Some 1.15 million people have been killed by firearms in the United States since Lennon was gunned down'', ``\#StopGunViolence'' \\ 
  \bottomrule
  \end{tabular}
\end{table*}

Social media have emerged as an important source of information for many worldwide. Unfortunately, not all information they publish is true.
During critical events such as a political election or a pandemic outbreak, disinformation with malicious intent~\cite{shu2017fake}, commonly known as ``fake news'', can disturb social behavior, public fairness, and rationality. As part of the fight against COVID-19, the World Health Organization also addressed the \textit{infodemic} caused by fatal disinformation related to infections and cures~\cite{thomas_2020}.

Many sites and social media have devoted efforts to identify disinformation. For example, Facebook encourages users to report non-credible posts
and employs professional fact-checkers to expose questionable news. Manual fact-checking is also used by fact-checking websites such as Snopes,
FactCheck,
PolitiFact,
and Full Fact.
In order to scale with the increasing amount of information, automated news verification systems consider external knowledge databases as evidence~\cite{hassan2017claimbuster,thorne2018factcheck,popat2018credeye}. Evidence-based approaches achieve high accuracy and offer potential explainability, but they also take considerable human effort.
Moreover, fact-checking approaches for textual claims based on textual evidence are not easily applicable to claims about images or videos.

Some recent work has taken another turn and has explored contextual features of the news dissemination process. 
They observed distinctive engagement patterns when social users face fake versus factual news~\cite{ma2016detecting,jin2016news}.
For example, the fake news shown in Table~\ref{table:temporal_engagement} had many engagements shortly after its publication. These are mainly verbatim
re-circulations with negative sentiment of the original post explained by the typically appalling content of fake news. After that short time window, we see denial 
posts questioning the validity of the news, and the stance distribution stabilizes afterwards with virtually no support. In contrast, the real news example in Table~\ref{table:temporal_engagement} invokes moderate engagement, mainly comprised of supportive posts with neutral sentiment that stabilize quickly. Such temporal shifts in user perception serve as important signals for distinguishing fake from real news. 

Previous work proposed partial representations of social context with (\emph{i})~news, sources and users as major entities, and (\emph{ii})~stances, friendship, and publication as major interactions~\cite{jin2014news,popat2017truth,shu2019beyond,popat2016credibility}. However, they did not put much emphasis on the quality of representation, modeling of entities and their interactions, and minimally supervised settings at all.  

\begin{figure}[t]
\centering
\includegraphics[scale=0.25]{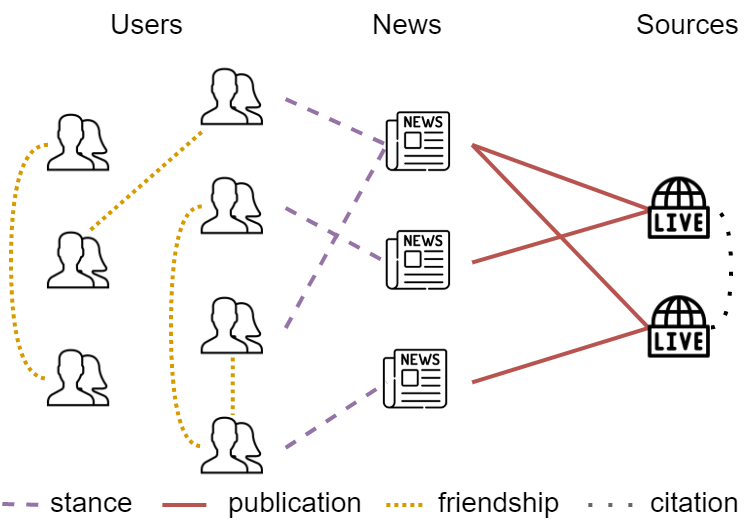}
\caption{Graph representation of social context.}
\label{fig:social_graph}
\end{figure}

Naturally, the social context of news dissemination can be represented as a heterogeneous network where nodes and edges represent the social entities and the interactions between them, respectively. 
Network representations have several advantages over some existing Euclidean-based methods~\cite{ruchansky2017csi,liu2018early} in terms of structural modeling capability for several phenomena such as
echo chambers of users or polarized networks of news media. Graphical models also allow entities to exchange information, via (\emph{i})~homogeneous edges, \textit{i.e.},~user--user relationship, source--source citations, (\emph{ii})~heterogeneous edges, \textit{i.e.},~user--news stance expression, source--news publication, as well as (\emph{iii})~high-order proximity (\textit{i.e.},~between users who consistently support or deny certain sources, as illustrated in Figure~\ref{fig:social_graph}). This allows the representation of heterogeneous entities to be dependent, leveraging not only fake news detection but also related social analysis tasks such as malicious user detection~\cite{SeminarUsers2017} and source factuality prediction \cite{baly2018predicting}.

Our work focuses on improving contextual fake news detection by enhancing representations of social entities. Our main contributions can be summarized as follows:
\begin{enumerate}
    \item We propose a novel graph representation that models all major social actors and their interactions (Figure~\ref{fig:social_graph}). 
    \item We propose the Factual News Graph (FANG), an inductive graph learning framework that effectively captures social structure and engagement patterns, thus improving representation quality.
    \item We report significant improvement in fake news detection 
    when using FANG and further show that our model is robust in the case of limited training data.
    \item We show that the representations learned by FANG generalize to related tasks such as predicting the factuality of reporting of a news medium.
    \item We demonstrate FANG's explainability thanks to the attention mechanism of its recurrent aggregator.
\end{enumerate}

\section{Related Work}\label{sec:related}
\begin{table*}[tbh]
  \centering
  \small
\caption{Comparison between representation learning frameworks for social entities (1. users, 2. news, 3. sources) and interactions (4. user-user friendship, 5. user--news engagement, 6. source--news publication, 7. source--source citation) on whether they consider engagement time, graph modelling of social context, deep learning, inductiveness, and representation learning.}
  \begin{tabular}{llccccc}
  \toprule
    \bf Approach & \bf Social Entities \& Interactions & \bf Temporal & \bf Graphical & \bf Deep & \bf Inductive & \bf Representative \\ 
    \midrule 
    Feature engineering~\cite{castillo2011information,ma2015detect,yang2012automatic,popat2016credibility} & 1, 2 & & & & \checkmark & \\
    Popat~\cite{popat2017truth} & 2, 3, 6 & \checkmark & & & & \\
    CSI~\cite{ruchansky2017csi} & 1, 2, 4, 5 & \checkmark & & \checkmark & \checkmark & \\
    TriFN~\cite{shu2019beyond} & 1, 2, 3, 4, 5, 6 & & \checkmark & & & \checkmark \\
    MVDAM~\cite{kulkarni2018multi} & 2, 3, 6, 7 & & \checkmark & \checkmark & & \\
    Monti~\cite{monti2019fake} & 1, 2, 4, 5 & \checkmark & \checkmark & & & \\
    GLAN~\cite{yuan2019jointly} & 1, 2, 5 & & \checkmark & \checkmark & & \\ \hline
    \textbf{FANG}~\tiny{(Our proposed approach)} & 1, 2, 3, 4, 5, 6, 7 & \checkmark & \checkmark & \checkmark & \checkmark & \checkmark \\ 
    \bottomrule
  \end{tabular}
  \label{table:literature_review}
\end{table*}

In this section, we first review the existing work on contextual fake news detection and the way the social context of news is represented in such work. We then detail recent advances in the Graph Neural Network (GNN) formalism, forming the premise of our proposed graph learning framework.

\subsection{Contextual Fake News Detection}
Previous work on contextual fake news detection can be categorized based on the approach used to represent and learn the social context. 

\emph{Euclidean approaches} represent the social context as a flat vector or a matrix of real numbers. 
They typically learn a Euclidean transformation of the social entity features that best approximates the fake news prediction~\cite{popat2016credibility}. The complexity of such transformation varies from the traditional shallow (as opposed to ``deep'') models, \textit{i.e.},~Random Forest or Support Vector Machines (SVM)~\cite{castillo2011information,yang2012automatic} to probabilistic graphical models~\cite{popat2017truth} and deep neural networks such as Long Short-Term Memory (LSTM)~\cite{lstm1997hochreiter} that model engagement temporality~\cite{ruchansky2017csi}. However, given our formulation of social context as a heterogeneous network, Euclidean representations are less expressive~\cite{bronstein2017geometric}. Although pioneering work used user attributes such as demographics, news preferences, and social features, \emph{e.g.}, the number of follower and friends~\cite{ma2015detect,shu2017fake}, they do not capture the user interaction landscape, \textit{i.e.},~what kind of social figures they follow, which news topics they favor or oppose, etc. Moreover, in graphical representation, node variables are no longer constrained by the independent and identically distributed assumption, and thus they can reinforce each other's representation via edge interactions.

Having acknowledged the above limitations, some researchers have started exploring non-Euclidean or \emph{geometric approaches}. They generalized the idea of using social context by modeling an underlying user or the news source network and by developing representations that capture structural features about the entity.

The \emph{Capture, Score, and Integrate} (CSI) model~\cite{ruchansky2017csi} used linear dimensionality reduction on the user co-sharing adjacency matrix, combining it with news engagement features obtained from a recurrent neural network (RNN). 

The \emph{Tri-Relationship Fake News} (TriFN) detection framework~\cite{shu2019beyond} -- although similar to our approach -- neither differentiated user engagements in terms of stance and temporal patterns, nor modeled source--source citations. Also, matrix decomposition approaches, including CSI~\cite{ruchansky2017csi}, can be expensive in terms of graph node counts and ineffective for modeling high-order proximity. 

Other work on citation source network~\cite{kulkarni2018multi}, propagation network~\cite{monti2019fake}, and rumor detection~\cite{yuan2019jointly,ming2019multiple} used recent advances in GNNs and multi-head attention to learn both local and global structural representations. 
These models optimized solely for the objective of fake news detection, without accounting for representation quality.
As a result, they are not robust 
when presented with limited training data and cannot be generalized to other downstream tasks, as we show in Section~\ref{sec:discussion}.
Table~\ref{table:literature_review} compares these approaches.

\begin{table*}[tbh]
  \centering
  \small
  \caption{Interactions in FANG's social context network.}
  \begin{tabular}{llllc}
  \toprule
    \bf Interaction & \bf Linking Entities & \bf Link Type & \bf Description & \bf Temporal \\ 
    \midrule
    Followership & User--user & Unweighted, undirected & Whether a user follows another user on social media & No \\
    Citation & Source--source & Unweighted, undirected & Whether sources refers to another source via a hyperlink & No \\
    Publication & Source--news & Unweighted, undirected & Whether the source published the target news & Yes \\
    Stance & User--news & Multi-label, undirected & The stance of the user with respect to the news & Yes \\ 
    \bottomrule
  \end{tabular}
  \label{table:social_interactions}
\end{table*}

\subsection{Graph Neural Networks (GNNs)}
GNNs have successfully generalized deep learning methods to model complex relationships and inter-dependencies on graphs and manifolds. Graph Convolutional Networks (GCNs) are among the first methods that effectively approximate convolutional filters~\cite{kipf2016semi}. However, GCNs impose a substantial memory footprint in storing the entire adjacency matrix.  They are also not easily adaptable to our heterogeneous graph, where nodes and edges with different labels exhibit different information propagation patterns. Furthermore, GCNs do not guarantee generalizable representations, and are transductive, requiring the inferred nodes to be present at training time. This is especially challenging for contextual fake new detection or general social network analysis, as their structure is constantly evolving.

With these points in mind, we build our work on GraphSage that generates embeddings by sampling and aggregating features from a node's local neighborhood~\cite{Hamilton2017InductiveRL}. 
GraphSage provides significant flexibility in defining the information propagation pattern with parameterized random walks and recurrent aggregators. It is well-suited for representation learning with unsupervised node proximity loss, and generalizes well in minimal supervision settings. Moreover, it uses a dynamic inductive algorithm that allows the creation of unseen nodes and edges at inference time.

\section{Methodology}

We first introduce the notation, and then formally define the problem of fake news detection. Afterwards, we discuss our methodology, namely the process of construction of our social context graph -- FANG -- as well as its underlying rationale. 
Finally, we describe the process of feature extraction from social entities as well as the modeling of their interactions. 

\subsection{Fake News Detection Using Social Context}
Let us first define the social context graph $G$ with 
its entities and interactions shown in Figure~\ref{fig:social_graph}:
\begin{enumerate}
    \item $A=\{a_1, a_2,...\}$ is the list of questionable \textbf{news articles}, 
    where each $a_{i}$ $(i=1,2,...)$ is modeled as a feature vector $\boldsymbol{x}_{a}$.
    \item $S=\{s_1, s_2,...\}$ is the list of \textbf{news sources}, where each source $s_j$ $(j=1,2,...)$ has published at least one article in $A$, and is modeled as a feature vector $\boldsymbol{x}_{s}$.
    \item $U=\{u_1, u_2,...\}$ is the list of \textbf{social users}, where each user $u_{k}$ $(k=1,2,...)$ has engaged in spreading an article in $A$, or is connected with another user; $u_{k}$ is a feature vector $\boldsymbol{x}_{u}$.
    \item $E=\{e_1, e_2,...\}$ is the list of {\bf interactions}, and each interaction $e=\{v_1, v_2, t, x_e\}$ is modeled as a relation between two entities $v_1, v_2\in A\cap S\cap U$ at time $t$;
    $t$ is absent in time-insensitive interactions. The interaction type of $e$ is defined as the label $x_{e}$.
\end{enumerate}

Table~\ref{table:social_interactions} summarizes the characteristics of different types of interactions, both homogeneous and heterogeneous. Stances are special types of interaction, as they are not only characterized by edge labels and source/destination nodes, but also by temporality as shown in earlier examples in Table~\ref{table:temporal_engagement}. Recent work has highlighted the importance of incorporating temporality not only for fake news detection~\cite{ruchansky2017csi,ma2015detect}, but also for modeling online information dissemination~\cite{he2014predicting}. We use the following stance labels: \textit{neutral support}, \textit{negative support}, \textit{deny}, \textit{report}. The major \textit{support} and \textit{deny} stances are consistent with the prior work~(\textit{e.g.,}~\cite{mohtarami2018automatic}), 
whereas the two types of \textit{support} ---\textit{neutral support} and \textit{negative support}--- are based on reported correlation between news factuality and invoked sentiment~\cite{Ajao2019}. We assign the \emph{report} stance label to a user--news engagement when the user simply spreads the news article without expressing any opinion. 
Overall, we use stances to characterize news articles based on opinions about them as well as social users by their view on various news articles.

\begin{table*}[tbh]
  \centering
  \small
    \caption{Some examples from the stance-annotated dataset, all concerning the same event.}
  \begin{tabular}{lll}
  \toprule
    \bf Text & \bf Type & \bf Annotated stance \\ \midrule
  Greta Thunberg tops annual list of highest-paid Activists! & reference headline & - \\
  Greta Thunberg is the ‘Highest Paid Activist’ & related headline & support (neutral) \\
  No, Greta Thunberg not highest paid activist & related headline & deny \\
  Can't speak for the rest of 'em, but as far as I know, Greta's just a schoolgirl and has no source of income. & related post & deny \\
  The cover describes Greta Thunberg to be the highest paid activist in the world & related tweet & support (neutral) \\
  A very wealthy 16yo Fascist at that! & related post & support (negative) \\ \bottomrule
  \end{tabular}
  \label{table:stance_annotation}
\end{table*}

\begin{table}[t]
  \centering
  \small
  \caption{Statistics about our stance-annotated dataset.}
  \begin{tabular}{crrr}
  \toprule
     & \bf \# Samples & \bf \# Supports & \bf \# Denies \\ 
    \midrule
  Train & 2,089 & 931 & 1,158 \\
  Test & 438 & 207 & 231 \\ 
  \bottomrule
  \end{tabular}
  \label{table:stance_statistics}
\end{table}

We can now formally define our task as follows: 
\begin{definition}{\textit{Context-based fake news detection}}: Given a social context $G = (A,S,U,E)$ constructed from news articles $A$, news sources $S$, social users $U$, and social engagements $E$, context-based fake news detection is defined as the binary classification task to predict whether a news article $a\in A$ is fake or real, in other words,
$F_C : a \rightarrow \{0,1\}$ such that,
\[  F_C(a) = \left\{
\begin{array}{ll}
      0 & \textrm{if } a \textrm{ is a fake article} \\
      1 & \textrm{otherwise}. \\
\end{array} 
\right. \]
\end{definition}

\subsection{Graph Construction from Social Context}
\label{sec:graph_construction}

\textbf{News Articles.} Textual~\cite{castillo2011information,yang2012automatic,shu2019beyond,popat2018credeye} and visual~\cite{wang2018eann,khattar2019mvae} features have been widely used to model news article contents, either by feature extraction, unsupervised semantics encoding, or learned representation.
We use unsupervised textual representations as they are relatively efficient to construct and optimize.
For each article $a\in A$, we construct a 
TF.IDF~\cite{Salton83} vector 
from the text body of the article. 
We enrich the representation of news by weighting the pre-trained embeddings from GloVe~\cite{pennington2014glove} of each word with its TF.IDF value, forming a semantic vector. Finally, we concatenate the TF.IDF and semantic vector to form the news article feature vector $\boldsymbol{x}_a$.

\textbf{News Sources.} 
We focus on characterizing news media sources using the textual content of their websites~\cite{baly2018predicting,kulkarni2018multi}. Similarly to article representations, for each source $s$, we construct the source feature vector $\boldsymbol{x}_s$ as the concatenation of its TF.IDF vector and its semantic vector derived from 
the words in the \emph{Homepage} and the \emph{About Us} section, as some fake news websites openly declare their content to be satirical or sarcastic.

\textbf{Social Users.} Online users have been studied extensively as the main propagator of fake news and rumors in social media. As in Section~\ref{sec:related}, previous work~\cite{castillo2011information,yang2012automatic} used attributes such as demographics, information preferences, social activity, and network structure such as the number of followers or 
friends. 
Shu {\it et al.}~\cite{shu2019beyond} conducted feature analysis of user profiles and pointed out the importance of signals derived from profile description and timeline content. A text description such as ``\emph{American mom fed up with anti american leftists and corruption. I believe in US constitution, free enterprise, strong military and Donald Trump \#maga}'' strongly indicates the user's political bias and suggests the tendency to promote certain narratives. We calculate the user vector $x_u$ as the concatenation of a pair consisting of a TF.IDF vector and a semantic vector derived from the user profile's text description. 

\textbf{Social Interactions.} For each pair of social actors $(v_i, v_j)\in A\cap S\cap U$, we add an edge $e=\{v_i, v_j, t, x_e\}$ to the list of social interactions $E$ if they are linked via interaction type $x_e$. Specifically, for following, we examine whether user $u_i$ follows user $u_j$; for publication, we check whether news article $a_i$ was published by source $s_j$; for citation, we examine whether the \emph{Homepage} of source $s_i$ contains a hyperlink to source $s_j$. In the case of time-sensitive interactions, \textit{i.e.},~\textit{publication} and \textit{stance}, we record their relative timestamp with respect to the article's earliest time of publication.

\begin{figure}[t]
\centering
\includegraphics[scale=0.18]{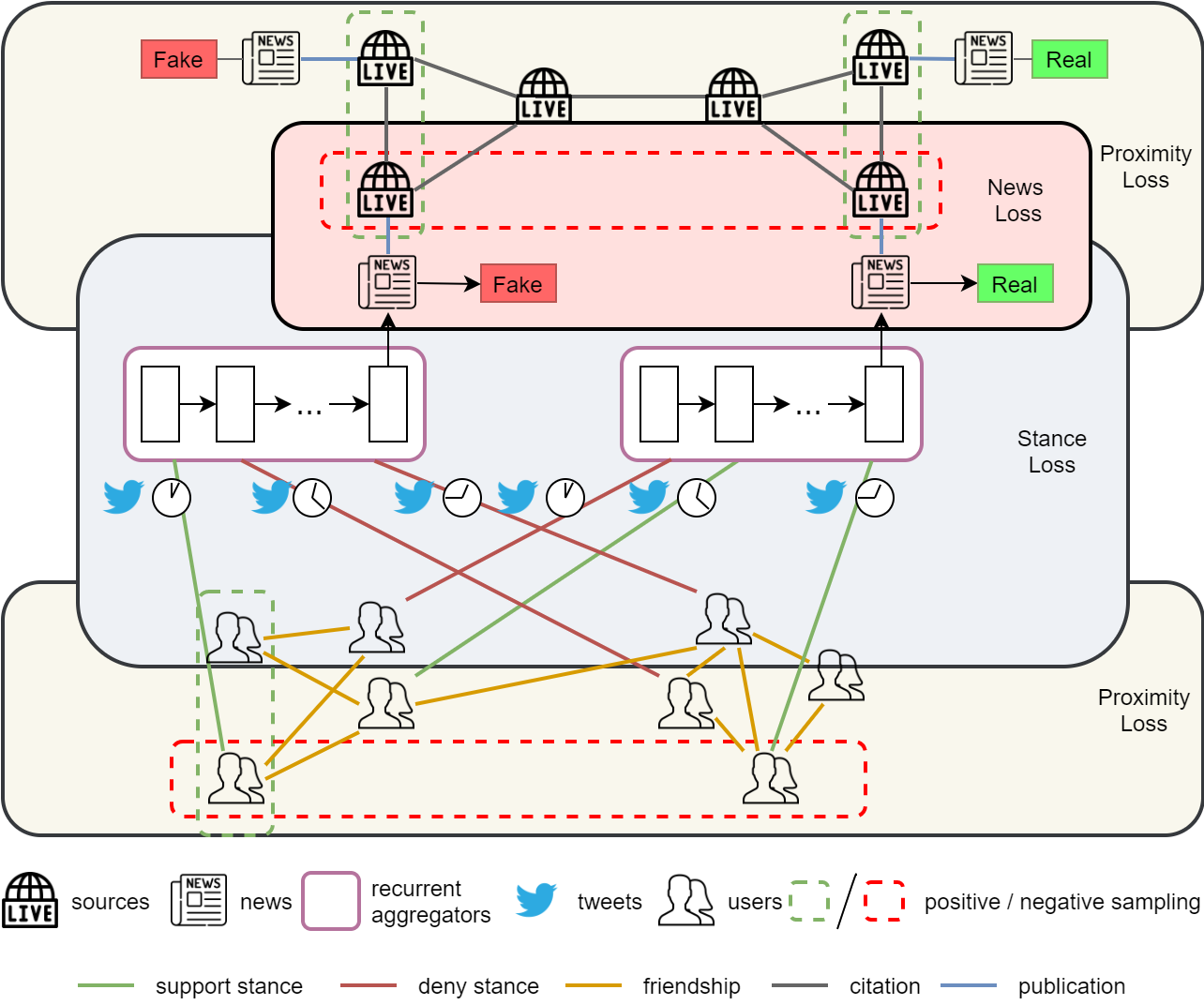}
\caption{Overview of our FANG framework.}
\label{fig:fang_overview}
\end{figure}

\textbf{Stance Detection.} The task of obtaining a viewpoint for a piece of text with respect to another one is known as \textit{stance detection}. 
In the context of fake news detection, we are interested in the stance of a user reply with respect to the title of a questionable news article.
We consider four stances: support with neutral sentiment or \textit{neutral support}, support with negative sentiment or \textit{negative support}, \textit{deny}, and \textit{report}. We classify a post as verbatim reporting of the news article if it matches the article title after cleaning the text from emojis, punctuation, stop words, and URLs.
We train a stance detector to classify the remaining posts as 
\textit{support} or \textit{deny}.
Popular stance detection datasets either do not explicitly describe the target text~\cite{derczynski-etal-2017-semeval}, have a limited number of targets~\cite{sobhani-etal-2017-dataset,mohammad-etal-2016-semeval}, or define the source/target texts differently, as in the \emph{Fake News Challenge}.\footnote{\scriptsize{\url{http://www.fakenewschallenge.org/}}} 

In order to overcome this difficulty, we constructed our own dataset for stance detection between social media posts and news articles, which contains 2,527 labeled source--target sentence pairs from 31 news events. For each event with a reference headline, the annotators were given a list of related headlines and posts. They labeled whether each related headline or post supports or denies the claim made by the reference headline. Aside from the \emph{reference headline--related headline} or the \emph{headline--related post} sentence pairs, we further made second-order inferences for \emph{related headline--related post} sentence pairs. If such a pair expressed a similar stance with respect to the reference headline, we inferred a {\it support} stance for the \emph{related headline--related post}, and \textit{deny}, otherwise. Tables~\ref{table:stance_annotation} and~\ref{table:stance_statistics} show example annotations and statistics about the dataset. The inter-annotator agreement evaluated with Cohen's Kappa is 0.78, indicating a substantial agreement. In order to choose the best stance classifier, we fine-tuned the model on our dataset using various pre-trained large-scale Transformers~\cite{devlin2019bert,liu2019roberta}. RoBERTa~\cite{liu2019roberta} turned out to work best, achieving \textit{Accuracy} of 0.8857, \textit{$F_1$ score} of 0.8379, \textit{Precision} of 0.8365, and \textit{Recall} of 0.8395, 
and thus we chose it for our stance classifier.

In order to further classify \textit{support} posts into such with neutral and with negative sentiment, we fine-tuned a similar architecture on the Yelp Review Polarity dataset to obtain a sentiment classifier. Altogether, the stance prediction of a user--article engagement $e$ is given as $stance(e)$.

\subsection{Factual News Graph (FANG) Framework}\label{sec:fang}

We now describe our FANG learning framework on the social context graph described in Section~\ref{sec:graph_construction}. Figure~\ref{fig:fang_overview} shows an overview of FANG.
While optimizing for the fake news detection objective, FANG also learns generalizable representations for the social entities. This is achieved by optimizing three concurrent losses: (\emph{i})~unsupervised \textit{Proximity Loss}, (\emph{ii})~self-supervised \textit{Stance Loss}, and (\emph{iii})~supervised \textit{Fake News Detection Loss}.

\textbf{Representation Learning}. We first discuss how FANG derives the representation of each social entity. Previous representation learning frameworks such as Deep Walk~\cite{perozzi2014deepwalk} and node2vec~\cite{grover2016node2vec} compute a node embedding by sampling its neighborhood, and then optimizing for the proximity loss similarly to word2vec. However, the neighborhood is defined by the graph structure. These methods use the neighborhood structure only, and they are suitable when the node auxiliary features are unavailable or incomplete, \emph{i.e.},~when optimizing for each entity's structural representation separately. Recently, GraphSage~\cite{Hamilton2017InductiveRL} was proposed to overcome this limitation by allowing auxiliary node features to be used jointly with proximity sampling as part of the representation learning. 

Let $GraphSage(\cdot)$ be GraphSage's node encoding function. Thus, we can now obtain the structural representation $\boldsymbol{z}_u\in \mathbb{R}^d$ of any user and source node $r$ as $\boldsymbol{z}_r=GraphSage(r)$, where $d$ is the structural embedding dimension. For news nodes, we further enrich their structural representation with user engagement temporality, which we showed to be distinctive for fake news detection in Section~\ref{sec:introduction} above. This can be formulated as learning an aggregation function $F(a,U)$ that maps a questionable news $a$, and its engaged users $U$ to a temporal representation $\boldsymbol{v}^{temp}_{a}$ that captures $a$'s engagement pattern. Therefore, the aggregating model (\textit{i.e.},~the aggregator) has to be time-sensitive. 
RNNs fulfill this requirement: specifically, the Bidirectional LSTM (Bi-LSTM) can capture a long-term dependency in information sequence in both the forward and the backward directions~\cite{lstm1997hochreiter}. On top of the Bi-LSTM, we further incorporate an attention mechanism that focuses on essential engagement during the encoding process. Attention is not only expected to improve the model quality but also its explainability~\cite{luong2015effective,devlin2019bert}. By examining the model's attention, we learn which social profiles influence the decision, mimicking human analytic capability. 

Our proposed LSTM input is a user--article engagement sequence $\{e_1, e_2,\cdots,e_{|U|}\}$. Let $meta(e_i)\in \mathbb{R}^l=(time(e_i),stance(e_i))$ be the concatenation of $e_i$'s elapsed time since the news publication and a one-hot stance vector. Each engagement $e_i$ has its representation $\boldsymbol{x}_{e_i}= (\boldsymbol{z}_{U_i}, meta(e_i))$, where $\boldsymbol{z_{U_i}}=GraphSage(U_i)$. A Bi-LSTM encodes the engagement sequence and outputs two sequences of hidden states: (\emph{i})~a forward sequence,
$H^f={\boldsymbol{h}^f_1, \boldsymbol{h}^f_2,\ldots,
  \boldsymbol{h}^f_n}$, which starts from the beginning of the engagement sequence, and
(\emph{ii})~a backward sequence, $H^b={\boldsymbol{h}^b_1,
  \boldsymbol{h}^b_2,\ldots, \boldsymbol{h}^b_n}$, which starts from the
end of the engagement sequence. 

Let $w_i$ be the attention weight paid by our Bi-LSTM encoder to the forward ($\boldsymbol{h}^f_i$) and to the backward ($\boldsymbol{h}^b_i$) hidden states. This attention should be derived from the similarity of the hidden state and the news features, \textit{i.e.,}~how relevant the engaging users are to the discussed content, and the particular time and stance of the engagement. Therefore, we formulate the attention weight $w_i$ as:
\begin{equation}
    w_i = \frac{exp(\boldsymbol{z}_a \mathbf{M}_e  \boldsymbol{h}_i + meta(e_i) \mathbf{M}_m)}{\sum^n_{j=1}exp(\boldsymbol{z}_a \mathbf{M}_e \boldsymbol{h}_j + meta(e_j) \mathbf{M}_m)}.
\end{equation}

\noindent where $l$ is the meta dimension, $e$ is the encoder dimension, and $\mathbf{M}_e\in \mathbb{R}^{d\times e}$ and $\mathbf{M}_m\in \mathbb{R}^{l\times 1}$ are the optimizable projection matrices for engagement and meta features shared across all engagements. $w_i$ is then used to compute the forward and the backward weighted feature vectors as $\boldsymbol{h}^f=\sum^n_i w_i \boldsymbol{h}^f_i$ and $\boldsymbol{h}^b=\sum^n_i w_i \boldsymbol{h}^b_i$, respectively.

Finally, we concatenated the forward and the backward vectors to obtain the temporal representation $\boldsymbol{v}^{temp}_{a}\in\mathbb{R}^{2e}$ for article $a$. By explicitly setting $2e=d$, we can combine the temporal and the structural representations of a news $a$ into a single representation: 
\begin{equation}\label{eq:news_rep}
    \boldsymbol{z}_a=\boldsymbol{v}^{temp}_{a} + GraphSage(a).     
\end{equation}

\textbf{Unsupervised Proximity Loss}. We derive the Proximity Loss from the hypothesis that closely connected social entities often behave similarly. This is motivated by the echo chamber phenomenon, where social entities tend to interact with other entities of common interest to reinforce and to promote their narrative. This echo chamber phenomenon encompasses inter-cited news media sources publishing news of similar content or factuality, as well as social friends expressing similar stance with respect to news article(s) of similar content. Therefore, FANG should assign such nearby entities to a set of proximal vectors in the embedding space. We also hypothesize that loosely-connected social entities often behave differently from our observation that social entities are highly polarized, especially in left--right politics~\cite{boxell2017internet}. FANG should enforce that the representations of these disparate entities are distinctive.

The social interactions that define the above characteristics the most are user--user friendship, source--source citation, and news--source publication. As these interactions are either (a)~between sources and news or (b)~between news, we divide the social context graph into two sub-graphs, namely \emph{news--source sub-graph} and \emph{user sub-graph}. 
Within each sub-graph $G^{\prime}$, we formulate the following Proximity Loss function:
\begin{equation}\label{eq:proximity_loss}
    \mathcal{L}_{prox} =  - \sum_{u\in G^{\prime}} \sum_{r_p\in P_r} log(\sigma(\boldsymbol{z}_r^\top \boldsymbol{z}_{r_p})) + Q \cdot \sum_{r_n\in N_r} log(\sigma(-\boldsymbol{z}_r^\top \boldsymbol{z}_{r_n})), 
\end{equation}
where $z_r\in \mathbb{R}^d$ is the representation of entity $r$, $P_r$ is the set of nearby nodes or \textit{positive set} of $r$, and $N_r$ is the set of disparate nodes or \textit{negative set} of $r$.
$P_{r}$ is obtained using our fixed-length
random walk, and $N_{r}$ is derived using negative sampling~\cite{Hamilton2017InductiveRL}. 

\textbf{Self-supervised Stance Loss}.
We also propose an analogous hypothesis for the user--news interaction, in terms of stance. If a user expresses a stance with respect to a news article, their respective representations should be close. For each stance $c$, we first learn a user projection function $\alpha_c(u) = \mathbf{A_c} z_u$ and a news article projection function $\beta_c(a) = \mathbf{B_c} z_a$ that map a node representation of $\mathbb{R}^d$ to a representation in the stance space $c$ of $\mathbb{R}^{d_c}$. Given a user $u$ and a news article $a$, we compute their similarity score in the
stance space $c$ as $\alpha(u)^\top \beta(a)$. If $u$ expresses stance $c$ with respect to $a$, we maximize this score, and we minimize it otherwise. This is the stance classification objective, optimized using the Stance Loss: 
\begin{equation}\label{eq:stance_loss}
    \mathcal{L}_{stance} = -\sum_{u,a}\sum_{c} y_{u,a,c} log(f(u,a,c)), 
\end{equation}
where $f(u,a,c) = \text{softmax}(\alpha_c(u)^\top \beta_c(a))$ and
\[  y_{u,a,c} = \left\{
\begin{array}{ll}
      1 & \textrm{if } u \textrm{ expresses stance } c \textrm{ over } a, \\
      0 & \textrm{otherwise}. \\
\end{array} 
\right. \]

\textbf{Supervised Fake News Loss}. We directly optimize the main learning objective of fake news detection via the supervised Fake News Loss. In order to predict whether article $a$ is false, we obtain its contextual representation as the concatenation of its representation and the structural representation of its source, {\it i.e.}, $\boldsymbol{v}_a = (\boldsymbol{z}_a, \boldsymbol{z}_s)$. 

This contextual representation is then input into a fully connected layer whose outputs are computed as $o_a = \mathbf{W}\boldsymbol{v}_a+b$, where $\boldsymbol{W}\in \mathbb{R}^{2d\times 1}$ and $b\in \mathbb{R}$ are the weights and the biases of the layer. The output value $o_a \in \mathbb{R}$
is finally passed through a sigmoid activation function $\sigma(\cdot)$, and trained using the following cross-entropy Fake News Loss $\mathcal{L}_{news}$, defined as follows:   
\begin{equation}\label{eq:news_loss}
    \mathcal{L}_{news}=\frac{1}{T}\sum_a\{{y}_{a} \cdot log(\sigma({o}_{a}))+(1-{y}_{a})\cdot log(1-\sigma({o}_{a}))\}, 
\end{equation}
where $T$ is the batch size, $y_{a}=0$ if $a$ is a fake article, and $y_{a}=1$ otherwise.

We define the total loss by linearly combining these three component losses: $\mathcal{L}_{total} = \mathcal{L}_{prox.} + \mathcal{L}_{stance} + \mathcal{L}_{news}$. We provide detailed instructions for training FANG in Algorithm~\ref{algo:fang}.

\begin{algorithm}[t]
\SetAlgoLined
\SetKwInOut{Input}{Input}
\SetKwInOut{Output}{Output}
\Input{The social context graph $G = (A,S,U,E)$ \\
       The news labels $Y_{A}$, and the stance labels $Y_{U,A,C}$ \\
       }
\Output{FANG-optimized parameters $\theta$}
 Initialize $\theta$\;
 \While{$\theta$ has not converged}{
  \For{each news batch $A_i\subset A$}{
    \For{each news $a\in A_i$}{
        $U_a \leftarrow$ users who have engaged with $a$\;
        $z_a \leftarrow$ Equation~(\ref{eq:news_rep})\;
        $z_s \leftarrow GraphSage(s)$\;
        \For{each user $u\in U_a$}{
            $z_u \leftarrow GraphSage(u)$\;
            $\mathcal{L}^{\prime}_{stance} \leftarrow$ Equation~(\ref{eq:stance_loss})\;
        }
    }
    $\mathcal{L}^{\prime}_{news} \leftarrow$ Equation~(\ref{eq:news_loss})\;
  }
  \For{each news--source or user sub-graph $G^{\prime}$}{
    \For{each entity $r\in G^{\prime}$}{
        $P_r \leftarrow$ positive samples of $r$ in $G{^{\prime}}$\;
        $N_r \leftarrow$ negative samples of $r$ in $G{^{\prime}}$\;
        $\mathcal{L}^{\prime}_{prox.} \leftarrow$ Equation~(\ref{eq:proximity_loss})\;
    }
  }
  $\mathcal{L}_{total} \leftarrow$ SUM$(\mathcal{L}^{\prime}_{stance}, \mathcal{L}^{\prime}_{news}, \mathcal{L}^{\prime}_{prox.})$\;
  $\theta \leftarrow$ Backpropagate$(\mathcal{L}_{total})$\;
 }
 \Return{$\theta$}
 \caption{FANG Learning Algorithm}\label{algo:fang}
\end{algorithm}

\section{Experiments}

\subsection{Data}

We conducted our experiments on a Twitter dataset collected by related work on rumor classification~\cite{ma2016detecting, Kochkina2018} and fake news detection~\cite{shu2018fakenewsnet}. For each article, we collected its source, a list of engaged users, and their tweets if they were not already available in the previous dataset.
This dataset also
includes Twitter profile description and the list of Twitter profiles each user follows. 

We further crawled additional data about media sources, including the content of their \emph{Homepage} and their \emph{About us} page, together with their frequently cited sources on their \emph{Homepage}. 

The truth value of the articles, namely, whether they are fake or real news, is based on two fact-checking websites: 
Snopes and Politifact. We release the source code of FANG and the stance detection 
dataset.\footnote{\scriptsize{\url{https://github.com/nguyenvanhoang7398/FANG}}} Table~\ref{table:dataset_statistics} shows some statistics about our dataset. 

\subsection{Fake News Detection Results}\label{sec:macroscopic}
We benchmark the performance of FANG on fake news detection against several competitive models: (\emph{i})~a content-only model, (\emph{ii})~a Euclidean contextual model, and (\emph{iii})~another graph learning model. 
In order to compare our FANG with the content-only model, 
we use a Support Vector Machine (SVM) model on TF.IDF feature vectors constructed from the news content (see Section~\ref{sec:graph_construction}). 
We also compare our approach with a current Euclidean model,  
CSI~\cite{ruchansky2017csi}, a fundamental yet effective recurrent encoder that aggregates the user features, the news content, and the user--news engagements. We re-implement the CSI with source features by concatenating the overall score for the users and the article representation with our formulated source description to obtain
the result vector for CSI's Integrated module 
mentioned in the original paper. 
Lastly, we compare against the GCN graph learning framework~\cite{kipf2016semi}. First, we represent each of $k$ social interactions in a separated adjacency matrix. We then concatenate GCN's output on $k$ adjacency matrices as the final representation of each node, before passing the representation through a linear layer for classification.

We also verify the importance of modeling temporality by experimenting on two variants of CSI and FANG: 
(\emph{i})~temporal-insensitive CSI(-$t$) and FANG(-$t$) 
without $time(e)$ in the engagement $e$'s representation $\boldsymbol{x}_{e}$, and (\emph{ii})~temporal sensitive CSI and FANG with $time(e)$. 
Table~\ref{table:macroscopic} shows the macroscopic results. 
For evaluation, we use the area under the Receiver Operating Characteristic curve or AUC score as standard. 
All context-aware models (\textit{i.e.},~CSI(-$t$), CSI, GCN, and FANG(-$t$)) and FANG improve over the context-unaware baseline by 0.1153 absolute
with CSI(-$t$) and by 0.1993 absolute with FANG in terms of AUC score. This demonstrates that considering social context is helpful for fake news detection. %Secondly, 
We further observe that both time-sensitive CSI and FANG improve over their time-insensitive variants, CSI(-$t$) and FANG(-$t$) by 0.0233 and 0.0339, respectively. These results demonstrate the importance of modeling the temporality of news spreading. Finally, two graph-based models, FANG(-$t$) and GCN are consistently better than the Euclidean CSI(-$t$) 
by 0.0501 and 0.0386, respectively: this demonstrates the effectiveness of our social graph representation described in Section~\ref{sec:graph_construction}. Overall, we can observe that FANG outperforms the other context-aware, temporally-aware, and graph-based models.
\begin{table}[tb]
    \centering
    \small
    \caption{Statistics about our dataset.}
    \begin{tabular}{l@{}rlrlr} 
    \toprule
        Fake & 448 & Publications / source & 2.38 &  Cites / source & 8.38 \\
        Real & 606 & Engagements / news & 71.9  &  Friends / user & 58.25 \\
        Sources & 442 & Neu. support / news & 19.07 & Deny / news & 5.27 \\
        Users & 54461 & Neg. support / news & 10.83 & Report / news & 36.73 \\ 
        \bottomrule
    \end{tabular}
    \label{table:dataset_statistics}
\end{table}

\begin{table}[tb]
    \centering
    \small
    \caption{Comparison between FANG and baseline models on fake news detection, evaluated with AUC score.}
    \begin{tabular}{l@{}cccc} 
    \toprule
        \bf Model & \bf Contextual & \bf Temporal & \bf Graphical & \bf AUC \\ 
    \midrule
        Feature SVM & & & & 0.5525 \\
        CSI(-$t$)~\tiny{(without $time(e)$)} & \checkmark & & &  0.6678 \\
        CSI & \checkmark & \checkmark & & 0.6911 \\ 
        GCN & \checkmark & & \checkmark & 0.7064 \\
        FANG(-$t$)~\tiny{(without $time(e)$)} & \checkmark & & \checkmark & 0.7179 \\ \hline
        \textbf{FANG} & \checkmark & \checkmark & \checkmark &  \textbf{0.7518} \\ 
        \bottomrule
    \end{tabular}
    \label{table:macroscopic}
\end{table}

\section{Discussion}\label{sec:discussion}

We now answer the following research questions (RQs) to better understand FANG's performance under different scenarios: 
\begin{itemize}
    \item RQ1: Does FANG work well with limited training data?
    \item RQ2: Does FANG differentiate between fake and real news based on their contrastive engagement temporality?
    \item RQ3: How effective is FANG's representation learning?
\end{itemize}

\subsection{Limited Training Data (RQ1)}

In order to address RQ1, we conducted the experiments described in Section~\ref{sec:macroscopic} using different sizes of the training dataset. We found consistent improvements over the baselines under both limited and sufficient data conditions.
Table~\ref{table:limited_data} shows the experimental results and Figure~\ref{fig:limited_and_attention} (left) further visualizes them. 
We can see that FANG consistently outperforms the two baselines for all training sizes: 10\%, 30\%, 50\%, 70\%, and 90\% of the data. In terms of AUC score at decreasing training size, among the graph-based models, GCN's performance drops by 16.22\% from 0.7064 at 90\% to 
0.5918 at 10\%, while FANG's performance drops by 11.11\% from 0.7518 at 90\% to 0.6683 at 10\%. We further observe that CSI's performance drops the least by only 7.93\% from 0.6911 at 90\% to 0.6363 at 10\%. 
Another result from an ablated baseline, FANG(-$s$), where we removed the stance loss, highlights the importance of this self-supervised objective. When training on 90\% of the data, the relative underperforming margin of FANG(-$s$) compared to FANG is only 1.42\% in terms of AUC. However, this relative margin increases as the availability of training data decreases, to at most 6.39\% at 30\% training data. Overall, the experimental results emphasize our model's effectiveness even at low training data availability compared to the ablated version, GNN and Euclidean, which confirms RQ1.

\subsection{Engagement Temporality Study (RQ2)} 

\begin{figure}[t]
\centering
\includegraphics[scale=0.25]{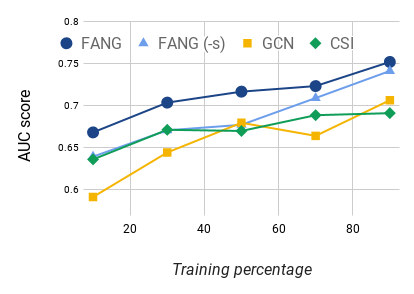}
\includegraphics[scale=0.35]{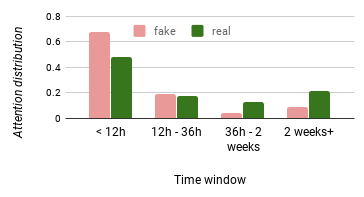}
\caption{FANG's performance against baselines (AUC score) for varying training data sizes (left), and attention distribution across time windows for fake vs. real news (right).}
\label{fig:limited_and_attention}
\end{figure}

To address RQ2 and to verify whether our model makes its decisions based on the distinctive temporal patterns between fake and real news, we examined FANG's attention mechanism. We accumulated the attention weights produced by FANG within each time window and we compared them across time windows. Figure~\ref{fig:limited_and_attention} (right) shows the attention distribution over time for fake and for real news. 
\begin{table}[t]
    \centering
    \small
    \caption{Comparison of FANG against baselines (AUC score) by varying the size of the training data.}
    \begin{tabular}{l  l l l l l} 
    \toprule
        \bf Systems & \multicolumn{5}{c}{\centering{\bf AUC score at different training percentages}} \\
         & \bf $10\%$ & \bf $30\%$ & \bf $50\%$ & \bf $70\%$ & \bf $90\%$ \\
    \midrule
        CSI & 0.6363  & 0.6714  & 0.6700 & 0.6887 & 0.6911 \\
        GCN & 0.5918 & 0.6445  & 0.6797 & 0.6642 & 0.7064 \\
        FANG(-$s$)~\tiny{(without stance loss)} & 0.6396 & 0.6708 & 0.6773 & 0.7090 & 0.7411 \\ \hline
         \textbf{FANG} &  \textbf{0.6683} &  \textbf{0.7036} &  \textbf{0.7166} &  \textbf{0.7232} &  \textbf{0.7518} \\ 
    \bottomrule
    \end{tabular}
    \label{table:limited_data}
\end{table}

We can see that FANG pays 68.08\% of its attention to the user engagement that occurred in the first 12 hours after a news article has been published to decide whether it is fake. Its attention then sharply decreases to 18.83\% for the next 24 hours, then to 4.14\% from 36 hours to two weeks after publication, and finally to approximately 9.04\% from the second week onward. 

On the other hand, for real news, FANG places only 48.01\% of its attention on the first 12 hours, which then decreases to 17.59\% and to 12.85\% in the time windows of 12 to 36 hours and 36 hours to two weeks, respectively. We also observe that FANG maintains 21.53\% attention even when the real news has been published after two weeks.

Our model's characteristics are consistent with the general observation that the appalling nature of fake news generates the most engagements within a short period of time after its publication. Therefore, it is reasonable that the model places much emphasis on these crucial engagements. On the other hand, genuine news attracts fewer engagements, but it is circulated for a longer period of time, which explains FANG's persistent attention even after two weeks after publication. Overall, the temporality study here highlights the transparency of our model's decision, largely thanks to the incorporated attention mechanism. 

\subsection{Representation Learning (RQ3)}

Our core claim is to improve the quality of representation with FANG, and we verify it in intrinsic and extrinsic evaluations. 

\begin{figure}[t]
\centering
\fbox{\includegraphics[scale=0.18]{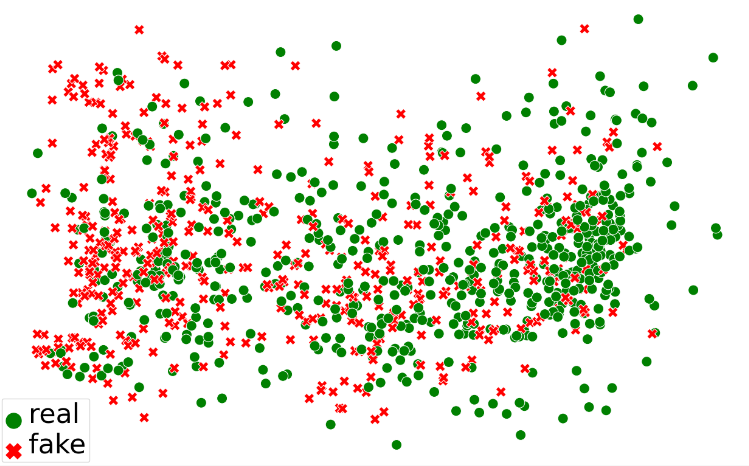}}
\fbox{\includegraphics[scale=0.18]{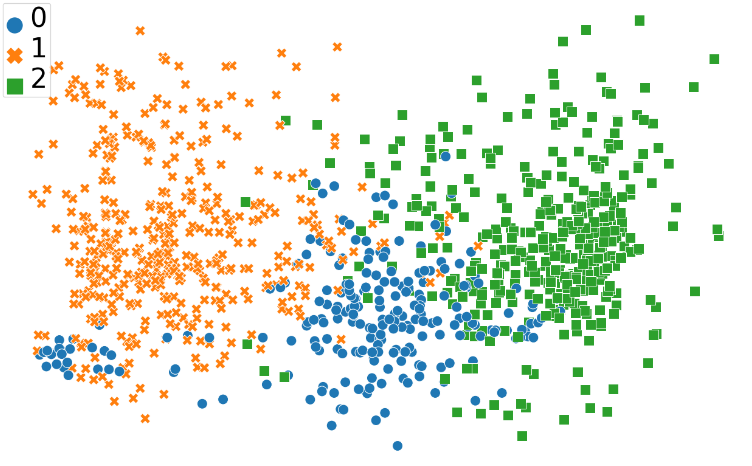}}\\
\fbox{\includegraphics[scale=0.18]{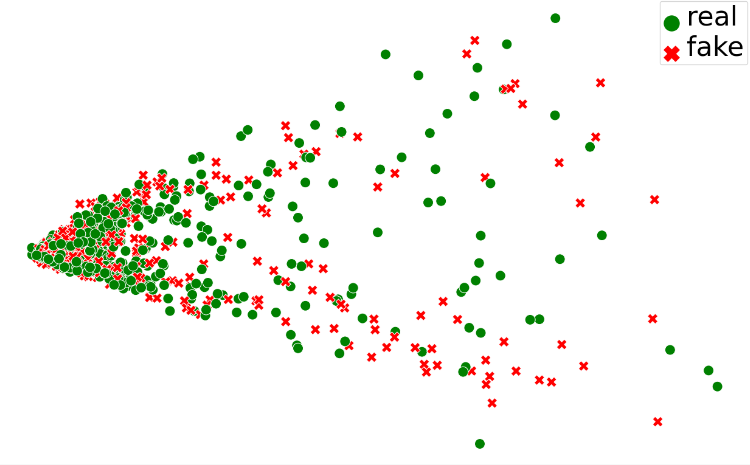}}
\fbox{\includegraphics[scale=0.18]{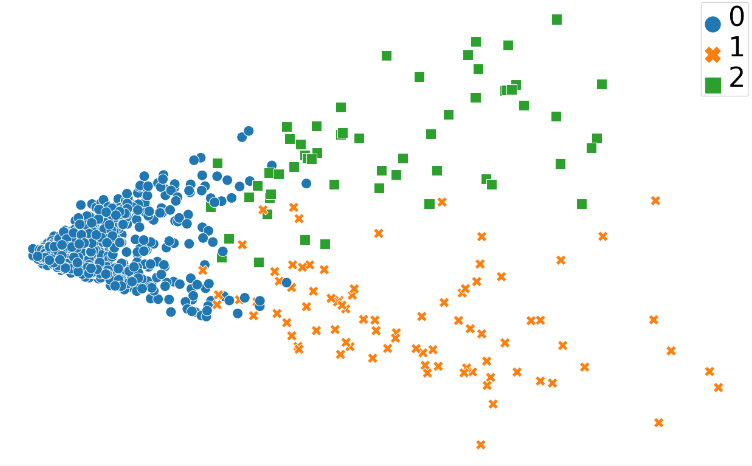}}\\
\caption{2D PCA plot of FANG's representations with factuality labels (top left) and OPTICS clustering labels (top right), and GCN's news representations with factuality labels (bottom left) and OPTICS clustering labels (bottom right).}
\label{fig:fang_rep}
\end{figure}

In the intrinsic evaluation, we verify how generalizable the minimally supervised news representations are for the fake news detection task. We first optimize both GCN and FANG on 30\% of the training data to obtain news representations. We then cluster these representations using an unsupervised clustering algorithm, OPTICS~\cite{ankerst1999optics}, and we measure the homogeneity score --- the extent to which clusters contain a single class. 
The higher the homogeneity score, the more likely the news articles of the same factuality label (\textit{i.e.},~fake or real) are to be close to each other, 
which yields higher quality representation. Figure~\ref{fig:fang_rep} visualizes the representations obtained from two approaches with factuality labels and OPTICS clustering labels. 

In the extrinsic evaluation, we verify how generalizable the supervised source representations are for a new task: source factuality prediction. We first train FANG on 90\% of the training data to obtain all source $s$ representations as $z_s=GraphSage(s)$, and the total representation as $v_s=(z_s,x_s,\sum_{a\in publish(s)}x_a)$, where $x_s$, $publish(s)$, and $x_a$ denote the source $s$ content representation, the list of all articles published by $s$, and their content representations. 

We propose two baseline representations that do not consider the source $s$ content, $v^{\prime}_s=(z_s,x_s)$. Finally, we train two separate SVM models for $v_s$ and $v^{\prime}_s$ on the source factuality dataset, consisting of 129 sources of high factuality and 103 sources of low factuality,
obtained from Media Bias/Fact Check\footnote{\scriptsize{\url{https://www.mediabiasfactcheck.com}}} 
and PolitiFact.\footnote{\scriptsize{\url{http://politifact.com}}}

\begin{figure*}[t]
\centering
\fbox{\includegraphics[width=5.5cm, height=4.5cm]{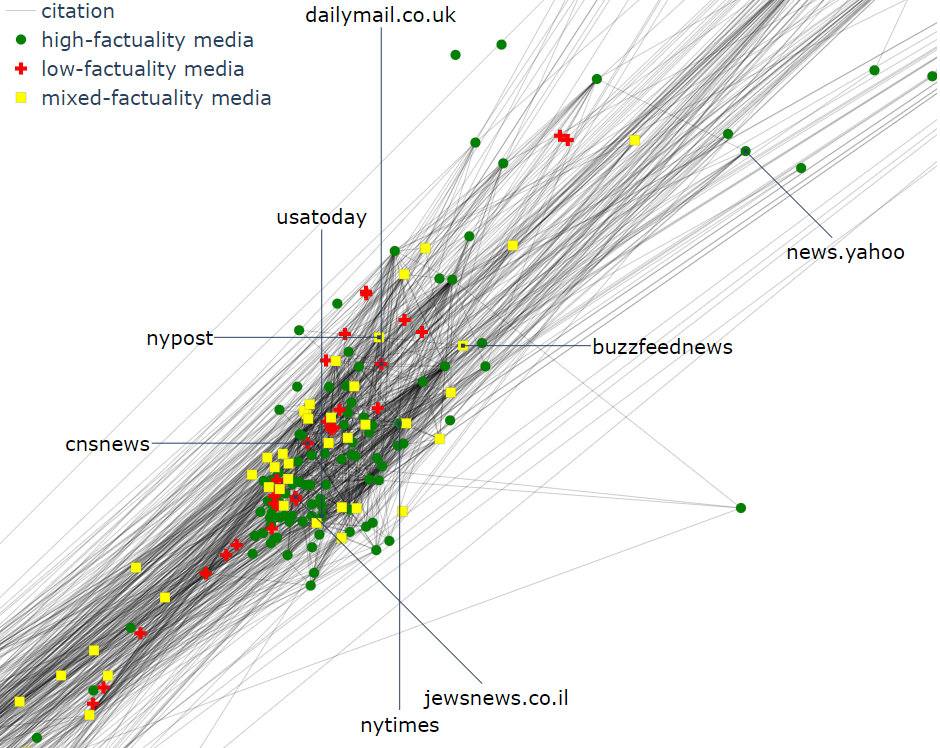}}
\fbox{\includegraphics[width=5.5cm, height=4.5cm]{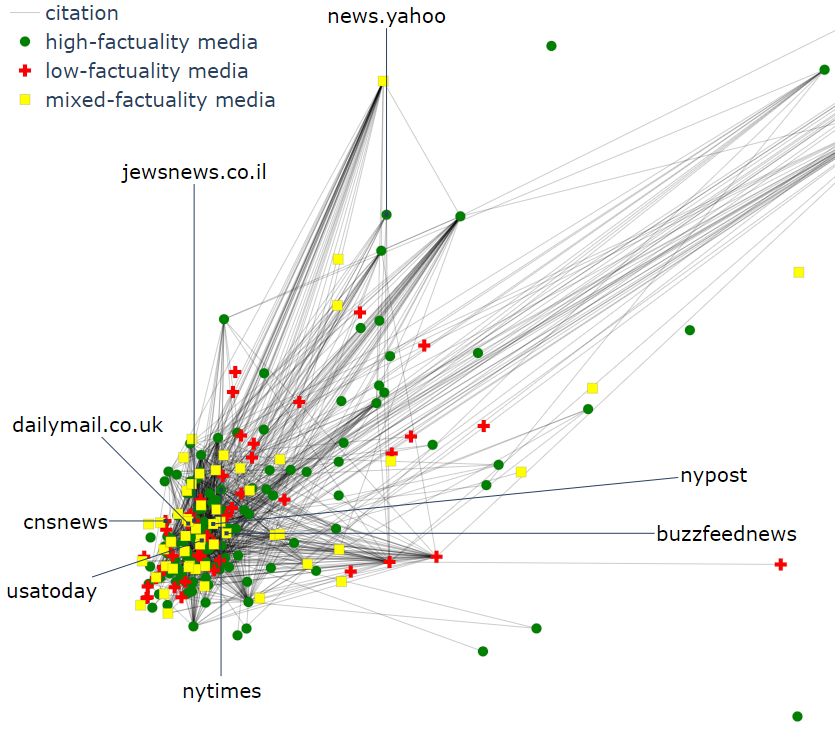}}
\fbox{\includegraphics[width=5.5cm, height=4.5cm]{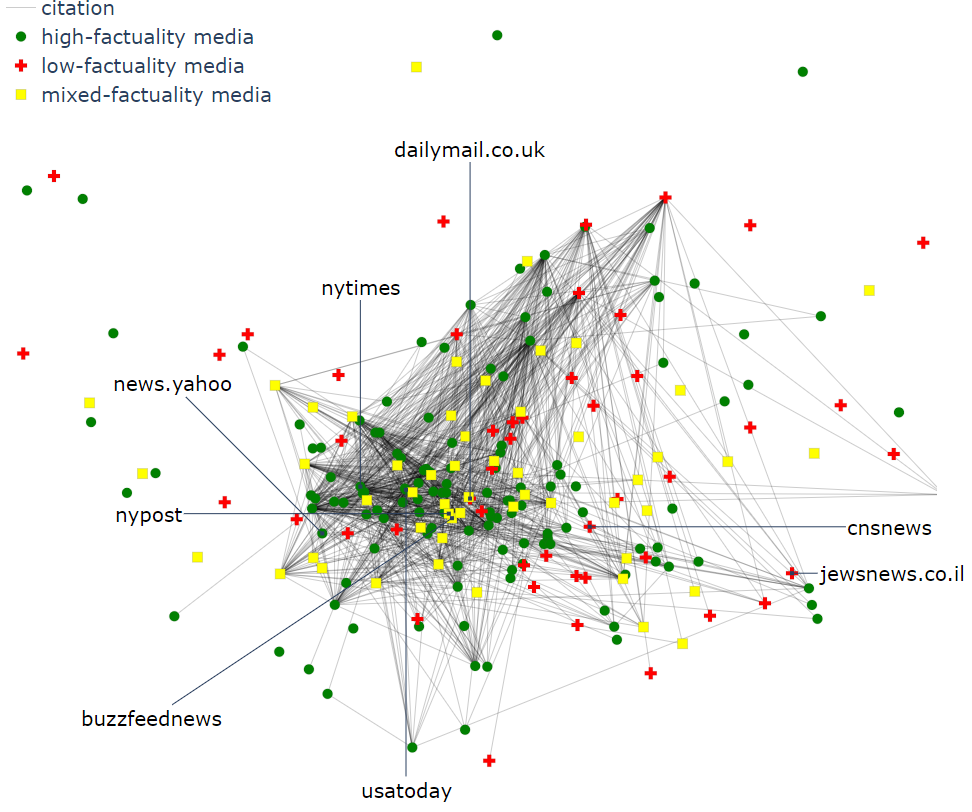}}\\
\caption{Plots for source representations using textual features (left), GCN (middle), and FANG (right) with factuality labels.}
\label{fig:source_rep}
\end{figure*}

For intrinsic evaluation, the Principal Component Analysis (PCA) plot of labeled FANG representation (see Figure~\ref{fig:fang_rep} top left) shows moderate collocation for the groups of fake and real news, while the PCA plot of labeled GCN representation (Figure~\ref{fig:fang_rep} bottom left) shows little collocation within either the fake or the real news groups. Quantitatively, FANG's OPTICS clusters (shown in Figure~\ref{fig:fang_rep} top right) achieve a homogeneity score of 0.051 based on news factuality labels, compared to 0.0006 homogeneity score for the GCN OPTICS clusters. This intrinsic evaluation demonstrates FANG's strong representation closeness within both the fake and the real news groups, indicating that FANG yields improved representation learning over another fully supervised graph neural framework. 

For the extrinsic evaluation on downstream source factuality classification, our context-aware model achieves an AUC score of 0.8049 compared to 0.5842 for the baseline. We further examined the FANG representations for sources to explain this 0.2207 absolute improvement. Figure~\ref{fig:source_rep} shows the source representations obtained from the textual features, GCN, and FANG with their factuality labels,  \textit{i.e.},~high, low, mixed, and citation relationship. In the left sub-figure, we can observe that the textual features are insufficient to differentiate the factuality of media, as a fake news spreading site such as \textit{cnsnews}
could mimic factual media in terms of web design and news content.

However, the citation between a low-factuality website and high-factuality sites would not be as high, and it is effectively used by the two graph learning frameworks: GCN and especially FANG. 
Yet, GCN fails to differentiate low-factuality sites with higher citations, such as \textit{jewsnews.co.il} and \textit{cnsnews}, from high-factuality sites. On the other hand, sources such as \textit{news.yahoo}
despite being textually different, as shown in Figure~\ref{fig:source_rep} (left), should still cluster with other credible media for their high inter-citation frequency. FANG, with much more emphasis on contextual representation learning, makes these sources more distinguishable. Its representation space gives us a glance into the landscape of news media, where there is a large central cluster of high-factuality inter-cited sources such as \textit{nytimes},
\textit{washingtonpost}
and \textit{news.yahoo}. At the periphery lie less connected media including both high- and low-factuality ones.

We also see cases where all models failed to differentiate mixed-factuality media, such as \textit{buzzfeednews} and
\textit{nypost},
which have high citation counts with high-factuality media.
Overall, the results from intrinsic and extrinsic evaluation, as well as the observations, confirm RQ3 on the improvement of FANG's representation learning.

\subsection{Scalable Inductiveness}
FANG overcomes the transductive limitation of previous approaches while inferring the credibility of unseen nodes. MVDAM~\cite{kulkarni2018multi} has to randomly initialize an embedding and to optimize it iteratively using node2vec~\cite{grover2016node2vec} for any unseen node, whereas FANG directly infers the embedding with its learned feature aggregator. Other graphical approaches using matrix factorization~\cite{shu2019beyond} or graph convolutional layers~\cite{monti2019fake,ming2019multiple} learn parameters whose dimensionality is fixed to the network size $N$, and can be as expensive as $O(N^3)$~\cite{ming2019multiple} in terms of inference time complexity. FANG infers the embeddings of unseen nodes without reconstructing the adjacency matrix, and its inference time complexity only depends on the size of the neighborhood of unseen nodes.

\subsection{Microscopic Analysis}
\begin{figure}[t]
\centering
\includegraphics[scale=0.25]{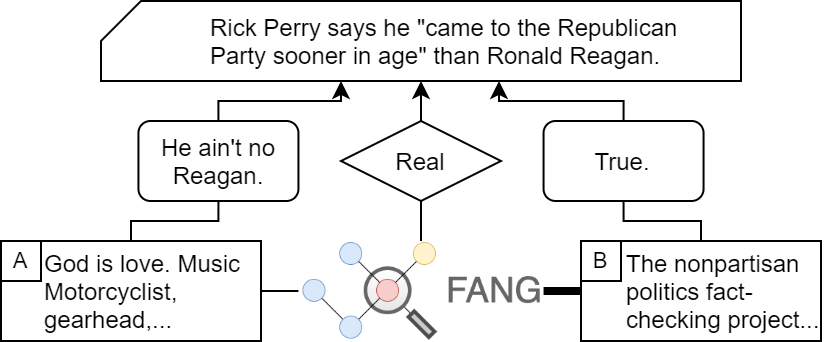}
\caption{A test example explaining FANG's decision.}
\label{fig:micro_1}
\end{figure}

It is also helpful to analyze FANG's predictions by examining specific test examples.  The first example is shown in Figure~\ref{fig:micro_1},
where we can see that FANG pays most of its attention to a tweet by \textit{user B}. This can be explained by B's Twitter profile description of a fact-checking organization, which indicates high reliability. 

In contrast, a denying tweet from \textit{user A} is not paid so much attention, due to the insignificant description of its author's profile. Our model bases its prediction on the \emph{support} stance from the fact-checker, which is indeed the correct label.

\begin{figure}[t]
\centering
\includegraphics[scale=0.25]{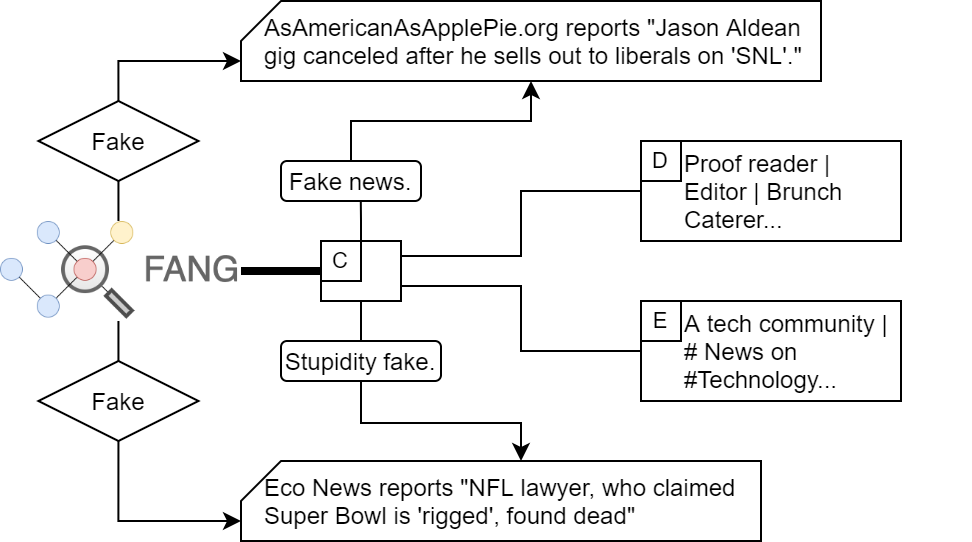}
\caption{Another test example explaining FANG's decision.}
\label{fig:micro_2}
\end{figure}

In the second example, shown in Figure~\ref{fig:micro_2}, 
FANG pays most attention to a tweet by \textit{user C}. Although this profile does not provide any description, it has a record of correctly denying the fake news about the dead NFL lawyer. Furthermore, the profiles that follow Twitter user C, namely \textit{user D} and \textit{user E}, have credible descriptions of 
a proof reader and of a tech community, respectively. 
This explains why our model bases its prediction of the news being fake thanks to the reliable denial, which is again the correct label.

\subsection{Limitations}
We note that entity and interaction features are constructed before passing to FANG, and thus errors from upstream tasks, such as textual encoding or stance detection, can propagate to FANG. Future work can address this in an end-to-end framework, where textual encoding~\cite{devlin2019bert} and stance detection can be jointly optimized.

Another limitation is that the dataset for contextual fake news detection can quickly become obsolete as hyperlinks and social media traces at the time of publication might no longer be retrievable.

\section{Conclusion and Future Work}
We have demonstrated the importance of modeling the social context for the task of fake news detection. We further proposed FANG, a graph learning framework that enhances representation quality by capturing the rich social interactions between users, articles, and media, thereby improving both fake news detection and source factuality prediction. 
We have demonstrated the efficiency of FANG with limited training data and its capability of capturing distinctive temporal patterns between fake and real news with a highly explainable attention mechanism. 
In future work, we plan more analysis of the representations of social users. We further plan to apply multi-task learning 
to jointly address the tasks of fake news detection, source factuality prediction, and echo chamber discovery.

\bibliographystyle{ACM-Reference-Format}
\bibliography{sample-base}

\appendix

\end{document}